%% file: main.tex
\pdfoutput=1  
\documentclass[10pt, conference]{IEEEtran}
\IEEEoverridecommandlockouts

\usepackage{cite}
\usepackage{amsmath,amssymb,amsfonts}
\usepackage{graphicx}
\usepackage{textcomp}
\usepackage{xcolor}
\usepackage{enumitem}
\usepackage{bbm}
\usepackage{dsfont}
\usepackage{siunitx}

\usepackage{rotating}

\usepackage[braket]{qcircuit}
\usepackage{lipsum}
\usepackage{lipsum}
\usepackage[switch]{lineno}

\usepackage{listings} 
\usepackage[braket]{qcircuit}

\usepackage{comment}
\usepackage{tikz}

\usepackage{algorithm}
\usepackage{algpseudocode}
\usepackage{nicefrac}

\newcommand{\qcrank}{\textbf{\texttt{QCrank}}}
\newcommand{\dpqa}{\textbf{\texttt{DPQA}}}
\newcommand{\eh}{\textbf{\texttt{EHands}}}

\begin{document}

\title{ Compilation of \qcrank\ Encoding Algorithm for a Dynamically Programmable
Qubit Array  Processor 
}

\author{
\IEEEauthorblockN{
Jan Balewski\IEEEauthorrefmark{1},  % balewski@lbl.gov
Wan-Hsuan Lin\IEEEauthorrefmark{3}, % wanhsuanlin@g.ucla.edu
Anupam Mitra\IEEEauthorrefmark{5}, % AnupamMitra@lbl.gov
Milan Kornja\v{c}a\IEEEauthorrefmark{2}, % mkornjaca@quera.com 
Stefan Ostermann\IEEEauthorrefmark{2}, % sostermann@quera.com
Pedro L. S. Lopes\IEEEauthorrefmark{2},\\ % plopes@quera.com
Daniel Bochen Tan\IEEEauthorrefmark{4}, % danieltan@g.harvard.edu
Jason Cong\IEEEauthorrefmark{3}, % cong@cs.ucla.edu
}
\IEEEauthorblockA{\IEEEauthorrefmark{1} National Energy Research Scientific Computing Center, Lawrence Berkeley National Laboratory, Berkeley, CA, USA }
\IEEEauthorblockA{\IEEEauthorrefmark{3} University of California, Los Angeles, Los Angeles, CA 90095, USA }
\IEEEauthorblockA{\IEEEauthorrefmark{2} QuEra Computing Inc., Boston, MA, USA }
\IEEEauthorblockA{\IEEEauthorrefmark{4} Department of Physics, Harvard University, Cambridge, MA 02138, USA }
\IEEEauthorblockA{\IEEEauthorrefmark{5} Applied Mathematics and Computational Research Division, Lawrence Berkeley National Laboratory, Berkeley, CA, USA }

\IEEEauthorblockA{Email: balewski@lbl.gov}
}

\maketitle

\begin{abstract}
 Algorithm and hardware-aware compilation co-design is essential  for the efficient deployment of near-term quantum programs. We present a compilation case-study implementing QCrank -- an efficient encoding protocol for storing sequenced real-valued classical data in a quantum state -- targeting  neutral atom-based Dynamically
Programmable Qubit Arrays (DPQAs). We show how key features of neutral-atom arrays such as high qubits count, operation parallelism, multi-zone architecture, and natively reconfigurable connectivity  can be used to inform effective algorithm deployment. We identify algorithmic and circuit features that signal opportunities to implement them in a hardware-efficient manner. To evaluate projected hardware performance, we define a realistic noise model for DPQAs using parameterized Pauli channels, implement it in Qiskit circuit simulators, and assess QCrank's accuracy for writing and reading back 24-320 real numbers into 6-20 qubits. We compare DPQA results with simulated performances of Quantinuum's H1-1E and with experimental results from IBM Fez, highlighting promising accuracy scaling for DPQAs.
\end{abstract}


%%%%%%%%%%%%%%%%%%%%%%%%%%%%%%%%%%%%%%%%%%%%%%%%%%%%%%
%%%%%%%%%%%%%%%%%%%%%%%%%%%%%%%%%%%%%%%%%%%%%%%%%%%%%%

%%%%%%%%%%%%%%%%%%%%%%%%%%%%%%%%%%%%%%%%%%%%%%%%%%%%%%
%%%%%%%%%%%%%%%%%%%%%%%%%%%%%%%%%%%%%%%%%%%%%%%%%%%%%%

\section{Introduction}

Neutral-atom technologies have emerged as a promising approach to gate-based quantum computing, with experimental systems now reaching hundreds of qubits and achieving competitive two-qubit gate fidelities of approximately 99.5\%~\cite{Bluvstein_2022,Evered2023,evered2025probingtopologicalmatterfermion}. Although no commercial devices have yet been offered on the cloud yet—leaving official performance parameters and error models to be disclosed by vendors at a later stage—several key features have already been identified as hallmarks of this architecture. Notably, these include (i) mid-circuit qubits connectivity reconfiguration~\cite{Bluvstein_2022}, (ii) the ability to create distinct functional zones~\cite{Bluvstein2024}, 
and (iii) a high degree of operational parallelization enabled by various strategies for the simultaneous addressing of multiple qubits via global laser beams~\cite{Evered2023}. These features are intimately connected, enabled and aided by quantum-coherent atom shuttling capabilities, which have inspired the designation of these devices as \emph{dynamically programmable qubit arrays} (\dpqa s). Unlike field programmable qubit arrays (FPQAs)~\cite{wurtz2023aquila,Satzinger2023}, which allow only static, pre-execution circuit configuration, DPQAs enable real-time, mid-circuit reconfiguration of qubit connectivity and operations.

The ability to physically transport atoms without loss of quantum coherence, combined with the geometric constraint that two-qubit gates can only be applied between atoms brought into close proximity, enables selective operations on specific qubit subsets within an operation zone using a single optical pulse. Individual qubit addressing has also been demonstrated~\cite{Knollmann:24}.

The benefits offered by these features, however, come with the challenge of navigating a complex compilation space~\cite{tan2024compiling, Tan2025, lin2024reuseawarecompilationzonedquantum, ruan2024powermoveoptimizingcompilationneutral, stade2024abstractmodelefficientrouting, wang2024atomiquequantumcompilerreconfigurable, stade2025optimalstatepreparationlogical}. Several critical questions need to be addressed: How can we decompose quantum circuits into layers that maximize the effective application of both single-qubit and two-qubit gates, while minimizing shuttling  of atoms? How many functional zones yield optimal performance? What qubits layout should these zones have? 

In this work we assume plausible \dpqa\ hardware constraints and introduce application-specific compilation strategies that address these challenges.
We illustrate the optimization process with a \qcrank\ implementation for sequenced data encoding~\cite{qcrank-nature}. 
We compare its projected accuracy on a simulated  \dpqa\ with results from Quantinuum's H1-1E trapped-ion emulator~\cite{QuantinuumH1Emulator} and IBM Heron superconducting devices~\cite{ibm_QPU}.

% = = = = = = = = = = = = = = = = 
\subsection{Gate-based neutral atoms processor }

For our neutral-atom \dpqa\  model, we adopt an architecture inspired by Rb atom-based  devices at Harvard~\cite{Bluvstein2024,evered2025probingtopologicalmatterfermion}, and related Gemini-class quantum computers by  QuEra~\cite{rodriguez2024experimentaldemonstrationlogicalmagic}. 

Contemporary neutral-atoms quantum computers deterministically load atoms into laser tweezer traps generated by spatial light modulators (SLMs), creating static trapping configurations that can be segmented into distinct zones. For instance, one zone may consist of a dense, regular square lattice for qubit storage, while another features one or more rows of closely paired trapping sites dedicated to entangling operations. Single-qubit gates can either be executed globally—using in-plane laser beams to address the Rb ground hyperfine manifold—thereby rotating every atom in the operation zone, or locally using an out-of-plane focused laser beam. Site-selected atoms can be transported between zones and rearranged using another type of tweezers, generated and steered by acousto-optical deflectors (AODs). AODs can move several atoms in parallel, provided their paths do not cross. While recent literature has reported strategies for mid-circuit atom-selective measurements~\cite{evered2025probingtopologicalmatterfermion,Atom_MCM}, our \dpqa\ model assumes global destructive measurement of all qubits. In the following, we discuss \qcrank\ circuit optimization and moves compilation assuming a single-zone architecture, leaving multi-zone layouts for future studies.

\begin{table}[htbp]
\vspace{-12pt}
\begin{center}
  \caption{\dpqa\ noise model implemented in Qiskit.}
\vspace{-8pt}
\resizebox{\columnwidth}{!}{%
\begin{tabular}{lccp{1.8cm}}
\hline 
\textbf{Operation} \rule{0pt}{8pt}& \textbf{Noise channel} & \textbf{Strength} & \textbf{Circuit symbol}\\
\hline
local 1q  U & depol. & $p=4e-3$  & \rule{0pt}{13pt}      \raisebox{0.25em}{\Qcircuit @C=0.5em  @R=0.2em @!R{& \qw & \gate{U} & \gate{\mathrm{lue}} & \qw \\   & \qw  \qw  & \qw  \qw  & \qw  & \qw  & \qw \\}}
\\
global 1q  U & depol. & $p=$4e-4  & \rule{0pt}{19pt}      \raisebox{0.25em}{\Qcircuit @C=0.5em  @R=0.2em @!R{& \qw & \gate{U} & \gate{\mathrm{gue}} & \qw \\ & \qw & \gate{U} & \gate{\mathrm{gue}} & \qw\\}}
\\
atom move & Pauli & [3e-5, 3e-5, 3e-3]$~^\dag$  &  \rule{0pt}{17pt}      \raisebox{0.25em}{\Qcircuit @C=0.5em {& \qw & \gate{\mathrm{mve}} & \qw & }}
\\
CZ-spectator & Pauli & [5e-4, 5e-4, 2.5e-3]$~^\dag$  & \rule{0pt}{15pt}      \raisebox{0.25em}{\Qcircuit @C=0.5em  @R=0.2em @!R{
& \qw &  \gate{\mathrm{spe}} &\qw  \\
& \qw & \ctrl{1} &  \qw \\ 
& \qw & \control \qw  &\qw  \\
}}
\\
global CZ & Pauli &   [1.5e-3, 1.5e-4]$~^\ddag$ & \rule{0pt}{15pt}   set existing CZ gate as noisy\vspace{4pt}
\\
 measurement & Pauli & [6e-3, 0, 0]$~^\dag$ &     {\Qcircuit @C=0.5em {& \qw & \gate{\mathrm{spam}} & \qw &\meter }}
\\
\hline
\end{tabular}
}
\label{tab:noisy_gates}
\end{center}
\vspace{-6pt}
\dag ) $[p_x,p_y,p_z]$, ~~~
\ddag)  two-qubit Pauli channel has 15 terms, see Methods. 1st value is for $p_{IZ}$,$p_{ZI}$,$p_{ZZ}$, 2nd value is for the remaining 12 terms.
\vspace{-4pt}
\end{table}

Table~\ref{tab:noisy_gates} summarizes the noise channels for the \dpqa{} considered in this work. The chosen values should be viewed as a current state-of-the-art baseline. 
To reflect both the potential for improvement and possible unaccounted errors, we also analyze the cases with noise reduced or increased by 30\% from the baseline, as discussed in the Methods section.
Below we discuss only the key features:
\begin{enumerate}[label=(\roman*)]
    \item Single-qubit gates can be implemented using either global or local laser beams. This results in two different noise channels $gue$ and $lue$, respectively. Global addressing results in higher gate fidelity compared to local addressing.
    \item Atom shuttling induces a move-error—modeled as a Pauli channel error ($mve$) per atom.
     This noise type is largely independent of the distance atoms travel and primarily scales with the number of times an atom is `touched' by the AOD tweezer. For current and near-term systems, the duration of atom shuttling is on the order of \qtyrange{10}{100}{\micro\second} and contributes significantly to the total circuit execution time.\footnote{Notice that gate times are actually much faster, in the order of \qty{200}{\nano\second}.} However, since the expected  long relaxation ($T_1$) and decoherence ($T_2$) times for the \dpqa\ exceed \qty{1}{\second}, the impact of movement-related delays on result fidelity is  orders of magnitude smaller than the channel errors listed in Table~\ref{tab:noisy_gates}. Therefore, we did not include the movement duration in this noise model.

    \item If qubits remain unpaired in the operations zone during a two-qubit operation, they incur a ``spectator" error ($spe$) since they still interact with the global laser beam.
    
    \item  Global measurement operation induces a standard Pauli-channel  error ($spam$).
\end{enumerate}

\subsection{\qcrank\ encoding and decoding}

We selected the \qcrank{} encoding algorithm to benchmark the expected performance of the noisy \dpqa{} on a practically relevant quantum computing tasks.
The \qcrank{} method~\cite{qcrank-nature} encodes real-valued data sequence onto data qubits through uniformly controlled rotation gates interspersed with entangling gates, thereby partially leveraging the exponential capacity of the Hilbert space.  This encoding is well suited for real-valued time series or grayscale images and is compatible with the \eh\ protocol, which enables vectorized polynomial operations on the \qcrank\ quantum state~\cite{balewski2025ehands}.

Generally, a \qcrank{} circuit with $n_a$ address qubits and $n_d$ data qubits can store $L = n_d \cdot 2^{n_a}$ real-valued numbers within the range $[-1,1]$, requiring $L$ single-qubit rotations around the $y$-axis ($\mathrm{Ry}$) and $L$ two-qubit entangling gates. In this study, we assume $n_d$ is always divisible by $n_a$, under which condition the circuit depth is $\nicefrac{n_d}{n_a} \cdot 2^{n_a } $ with a parallelism factor of $n_a$.

\begin{figure*}[htbp]
a) original \qcrank\ circuit\\
\vspace{-10pt}
\begin{center}
\includegraphics[width = 0.7\textwidth]{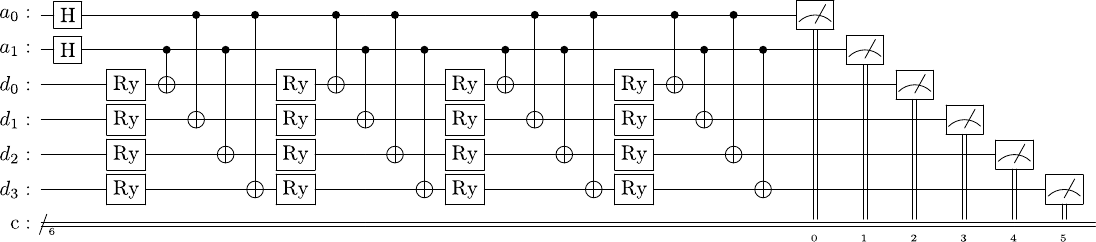}\\
\end{center}
b) optimized equivalent circuit\\
\vspace{-10pt}
\begin{center}
\includegraphics[width = 0.7\textwidth]{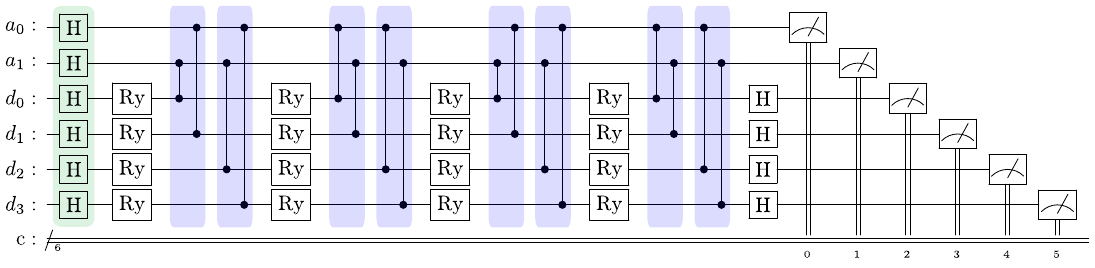}
\end{center}
\vspace{-12pt}
\caption{ \textbf{Full \qcrank\ circuit for 2 address ($\boldsymbol{a_i}$) and 4 data ($\boldsymbol{d_j}$) qubits} with  storage capacity of 16 real values. (a)  originally circuit proposed in Ref.~\cite{qcrank-nature},
(b) Equivalent circuit transpiled to a layout suited for neutral-atom hardware, featuring CZ parallelism of 2 and a single global Hadamard layer. Ry gates with various angles and some Hadamard gates on selected qubits are applied sequentially. Green-shaded blocks indicate global single-qubit gates, and purple-shaded blocks indicate groups of parallel CZ gates. This transpilation minimizes sequential single-qubit gates while maximizing global operations.
}
\vspace{-14pt}
\label{fig:qcrank2+4}
\end{figure*}

Figure~\ref{fig:qcrank2+4}(a) illustrates a complete \qcrank\ circuit example for $n_a=2$ and $n_d=4$, encoding 16 real values into $n_a+n_d=6$ qubits, with gates layout as originally proposed in~\cite{qcrank-nature}. 
Specifically, this circuit stores 4 sequences of 4 real numbers simultaneously.
Each data qubit labeled $d_0, \dots, d_3$ simultaneously stores 4 distinct values due to superposition with 2 address qubits. Data decoding involves measuring a selected data qubit and interpreting the result as the Z-basis expectation value. The two simultaneously measured address qubits specify the position within each sequence, taking values from $[0, 2^{n_a}-1]$, indicating which of the 4 values was just measured. A detailed description of the \qcrank\  encoding procedure is provided in~\cite{qcrank-nature}.

%%%%%%%%%%%%%%%%%%%%    CIRCUIT   %%%%%%%%%%%%%%%
 
%%%%%%%%%%%%%%%%%%%%    END OF CIRCUIT   %%%%%%%%%%%%%%%

 Transpiling the circuit from Fig.~\ref{fig:qcrank2+4}(a) to a neutral-atom native gate-set [arbitrary angle rotations and CZ entagling gates -- see Fig.~\ref{fig:qcrank2+4}(b)] reveals features and symmetries that are highly desired for \dpqa\ architectures. For one, the resulting quantum circuit exhibits a high degree of execution parallelism - understood here as \emph{identical} single-qubit gates acting on all qubits [highlighted in green in Fig.~\ref{fig:qcrank2+4}(b)] or entangling gates that can be applied in parallel layers across several qubits concurrently [highlighted in purple in Fig.~\ref{fig:qcrank2+4}(b)]. Furthermore, the ensuing bipartite connectivity between the address and data qubits is naturally well-suited for the \dpqa{} architecture, which inherently provides reconfigurable connectivity.

We quantify accuracy using the root mean square error (RMSE). 
To be precise, RMSE as our inaccuracy metric, reflects errors introduced by the noisy execution of a \qcrank{} circuit on the \dpqa.
Given the known input sequence encoded using \qcrank{} and the corresponding measurement outcomes from the QPU, we compute the RMSE between the ground-truth data and the measured data.

% !TEX root = 0-main.tex

%%%%%%%%%%%%%%%%%%%%    CIRCUIT   %%%%%%%%%%%%%%%
% JAN - DO NOT REMOVE
% cd ~/noisy_simu$ 
% ./noisy_qcrankV3.py -q 4 8  --mockCirc --skipNoisyOps  all -v3
% ./noisy_qcrankV3.py -q 4 8  --mockCirc  -v3

\begin{figure}[htbp]
\centering
\includegraphics[width=0.48\textwidth]{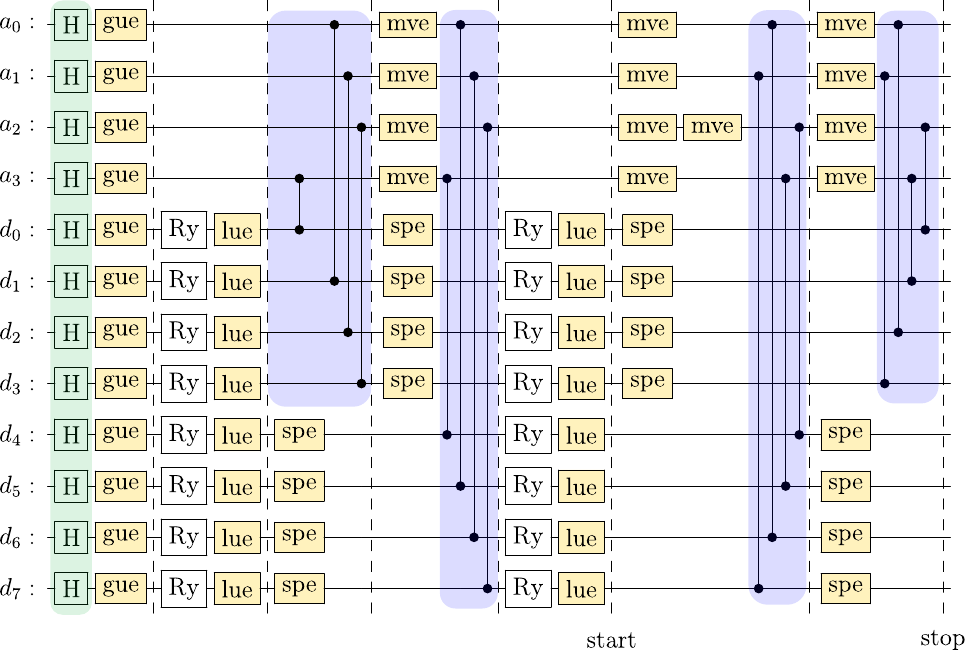}
\vspace{-5pt}
\caption{ \textbf{\qcrank\ circuit with 4 address and 8 data qubits including noise channels.}
Only the initial barrier-separated cycles are shown; the full circuit is deeper. Final measurements are omitted.
 Purple (green) shaded blocks indicate global two-qubit (single qubit) gates, leveraging the \dpqa{}'s native parallelism. Yellow boxes indicated  different noise channels detailed in Table~\ref{tab:noisy_gates}, added to simulate noise expected on \dpqa. The start/stop labels refer to the section of the circuit analyzed in more details in Fig.~\ref{fig:noZone-moves}.
}
\vspace{-16pt}
\label{fig:qcrank_4+8}
\end{figure}
%%%%%%%%%%%%%%%%%%%%    END OF CIRCUIT   %%%%%%%%%%%%%%%

\section{Results}

We describe  compilation optimization strategy by analyzing a more complex \qcrank\ configuration with $n_a = 4$ address qubits and $n_d = 8$ data qubits, encoding 128 real values onto 12 qubits. This setup achieves a CZ parallelism factor of 4 and a  CZ depth of 32. Due to the circuit's size, Fig.~\ref{fig:qcrank_4+8} shows only the initial cycles, which are sufficient to illustrate  our compilation approach.

%%%%%%%%%%%%%%%%%%%%
%%%%%%%%%%%%%%%%%%%%
\subsection{Atom movement strategy}
\begin{figure}[htb]
\centering
\includegraphics[width=0.35\textwidth]{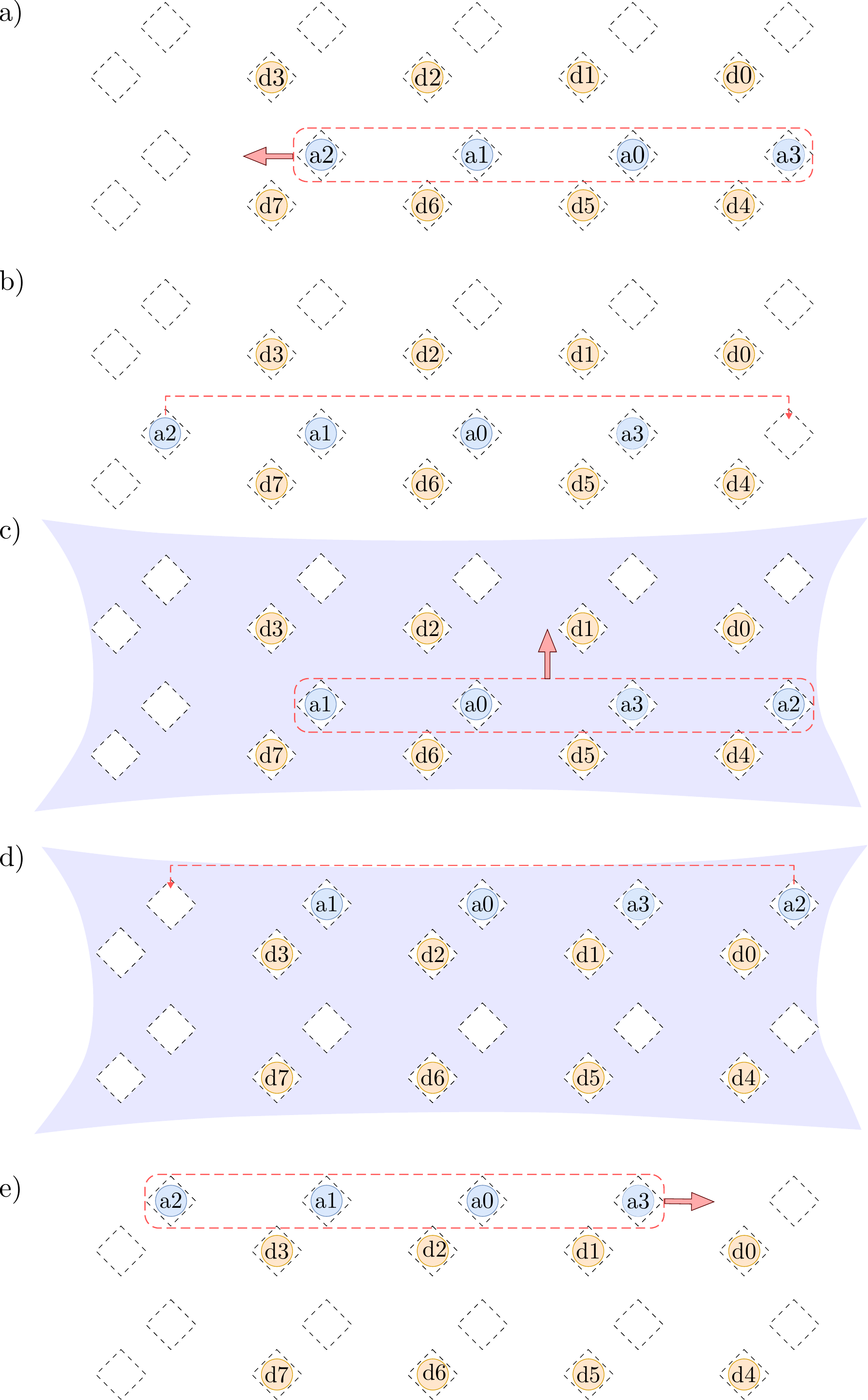}
\caption{
\textbf{Atoms placing  and movement patterns} for the circuit segment in Fig.~\ref{fig:qcrank_4+8} (between start \& stop barriers). White diamonds denote designated SLM spots where atoms can stay locked; 4 address (8 data) qubits are labeled with `a' (`d'); entangling pulses are denoted with purple shades. See text for the details.
}
\vspace{-18pt}
\label{fig:noZone-moves}
\end{figure}

 We adopt the rectangular qubits arrangement,  shown in Fig.~\ref{fig:qcrank_4+8}.
For optimal performance, we partition the data qubits into 2 sub-sets of size $n_a$, such that $d_0$--$d_{n_a-1}$ forms the first row and $d_{n_a}$--$d_{2n_a-1}$ the second. In each cycle, we apply CZ gates between address qubits and one sub-set of data qubits -  shown in Fig.~\ref{fig:noZone-moves}.
% Within each row, the qubits are ordered in decreasing index from left to right. 
Implementing a CZ gate sequence requires dynamically moving address qubits  into different data qubits  interaction  range.
 We follow three design principles: (i) only move address qubits, (ii) use horizontal moves for cyclic permutations between address qubits, and (iii) vertical moves for addressing different sets of data qubit.
To minimize vertical movement, we apply CZ gates with the first data row, then move the address qubits downward row by row. After reaching the last row, we reverse the process, moving upward to interact with each data group in reverse order.

Since the circuit requires cyclic qubits permutations, one might consider a circular layout with address qubits rotating around concentrically placed data qubits to realize CZ cycles. However, such an arrangement is incompatible with AOD movement constraints due to crossing atom paths.

Fig.~\ref{fig:noZone-moves} illustrates the execution for the circuit  shown in Fig.~\ref{fig:qcrank_4+8}, for the segment between the `start' and `stop' barriers. To adjust qubits pairings for the subsequent CZ cycle, we perform a cyclic shift of the four address qubits in two steps. First, all address qubits simultaneously move to the left, placing at least half directly into their target positions, as depicted in Fig.~\ref{fig:noZone-moves}(a).
In the second step, only a subset of address qubits moves in the opposite direction (Fig.~\ref{fig:noZone-moves}(b)), positioning them into their target locations, as shown in panel (c). This step completes the new mapping.
Subsequently, a global laser pulse induces CZ gates between the four qubit pairs $d_7$-$a_1$, \dots, $d_4$-$a_2$, indicated by the shaded region in panel (c). Since the number of data qubits in this example is twice the address qubits, data qubits are arranged across multiple rows. Address qubits are then shifted upward (still panel (c)), and a second global CZ pulse entangles the other four pairs $d_3$-$a_1$, \dots, $d_0$-$a_2$. At this point, the `stop' barrier is reached. Panel (e) illustrates how the next round of address qubit permutation begins from the upper row of the operation zone, thus further reducing the overall number of required atomic movements.

%%%%%%%%%%%%%%%%%%%%
%%%%%%%%%%%%%%%%%%%%
\subsection{Accuracy of \qcrank\ }
 Noise from both atoms movement and quantum gates was modeled using the Qiskit simulator with a density matrix backend. We evaluated \qcrank\ circuit performance across input sizes listed in Table~\ref{tab:qcrank_size} (Methods). To assess accuracy, we generated multiple sequences of random real numbers in $[-1,1]$, executed the circuits, and computed the RMSE.

\begin{figure}[htb]
\centering
\includegraphics[width=1.0\columnwidth]{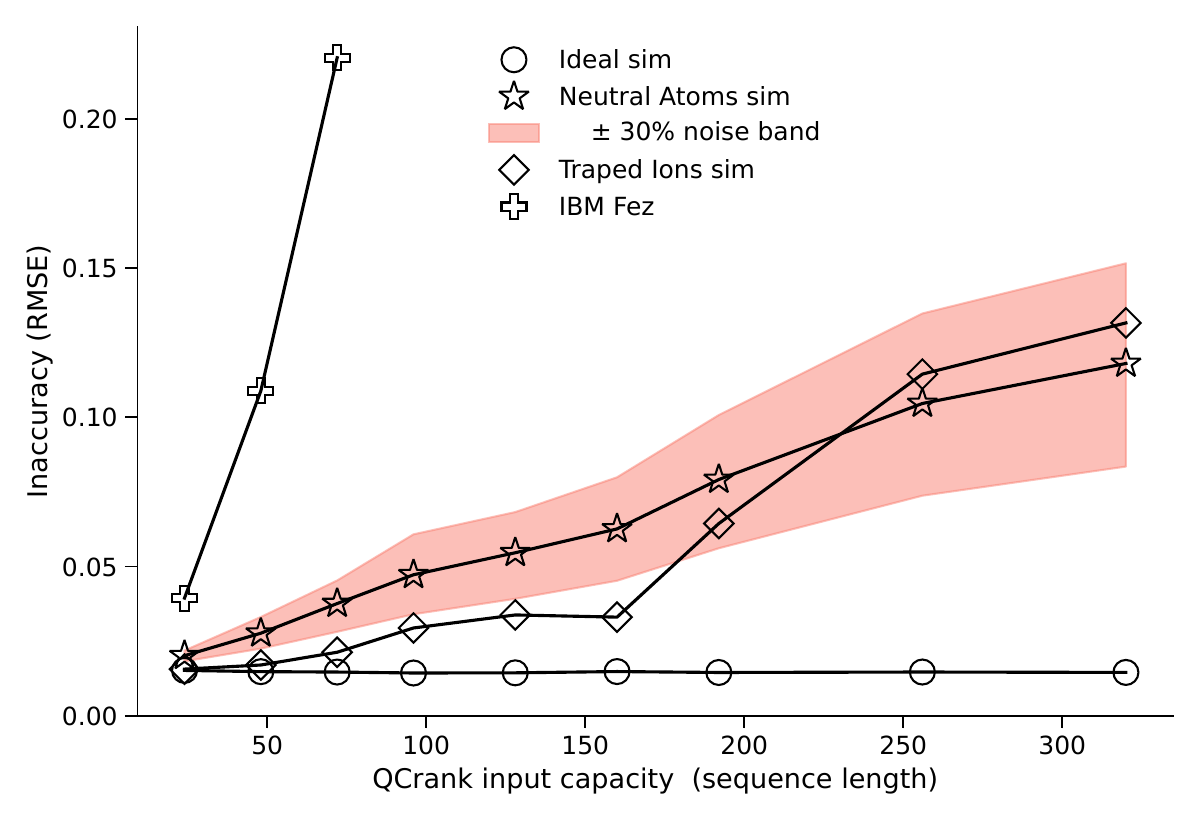}
\vspace{-24pt}
\caption{Dependence of inaccuracy of \qcrank\ vs. input size, for the same number of 3000 shots  per address. Simulated QPU backend types:   ideal  (circle);  neutral atoms   (stars) with salmon band  reflecting results obtained by changing the baseline noise level by $\pm30\%$; trapped ions (diamonds). Results from real IBM Fez QPU ar shown as crosses. 
}
\vspace{-8pt}
\label{fig:rmse_scaling}
\end{figure}

The main result of this work, shown as stars in Fig.~\ref{fig:rmse_scaling}, is the projected inaccuracy of \qcrank{} executed on future \dpqa{} devices. The longer input sequences require an increasing number of entangling gates, whose execution is the dominant source of inaccuracy. The salmon-colored band indicates the results obtained by scaling all noise baseline parameters (see Table~\ref{tab:noisy_gates}) by $\pm 30\%$. Tuning the noise level enables us to explore a broader range of potential performance scenarios for \dpqa{}, accounting for noise reduction in the near future and the possibility that certain errors may not be accurately represented by the average noise model.

All results  were assessed  using about 3000 shots per \qcrank{} address. The statistical residual RMSE of approximately 0.015 is constant, confirmed by  ideal Qiskit simulations,  illustrated by open circles. To contextualize our findings within the broader landscape of quantum processors, we evaluated identical circuits using the Quantinuum noisy simulation (H1-1E)~\cite{PhysRevX.13.041052,QuantinuumH1Emulator}, shown as diamonds in the figure. Note that no optimizations of the compilation strategy were performed beyond the default H1-1E simulator options.
These simulations suggest comparable expected \qcrank\ accuracy for digital neutral atom and trapped-ion QPUs across the evaluated sequence sizes. However, our simulations show discrepancies in the performance of different hardware platforms depending on the ratio between number of address and data qubits $\nicefrac{n_a}{n_d}$. If $\nicefrac{n_a}{n_d}=1$, as is the case in the 160-data input example with $n_d=n_a=5$, the 1D layout of the H1-1E processor exhibits superior performance (see Fig.~\ref{fig:rmse_scaling}). However, its relative performance decreases as $n_d$ grows relative to $n_a$ (compare Fig.~\ref{fig:rmse_scaling} and Table~\ref{tab:qcrank_size}). This non-uniformity is most likely related to the limited degree of parallelism for gate execution and qubit reconfiguration on Quantinuum's H1-1E chip~\cite{PhysRevX.13.041052}. Exploring alternative compilation strategies for race-track processors that could result in more uniform behavior across sequence lengths is a promising direction for future research.

To further enrich the analysis, we executed selected \qcrank{} circuits on IBM Fez with Pauli Twirling—currently the most effective error mitigation method available on this hardware~\cite{AbuGhanem_2025,IBMQuantumDocumentation_FakeFez}. Due to the limited native connectivity of IBM devices, transpiled circuits included additional entangling gates facilitating swaps  (Table~\ref{tab:qcrank_size}), resulting in larger RMSE values, which are  shown  as crosses in Fig.~\ref{fig:rmse_scaling}.

\section{Discussion}
Our results highlight several key design rules for the efficient compilation of quantum circuits for \dpqa s. We show how high-connectivity and parallellism can be leveraged, emphasizing the value of architectural flexibility provided by \dpqa. Hardware-aware compilation on the special example of the \qcrank~algorithm shows that using these key features result in the \dpqa's performance being on-par with alternative state-of-the-art quantum computing platforms, with promising scaling trends as qubit number increases (see Fig.~\ref{fig:rmse_scaling}).

Our exploration underlined important compiler design guidelines, particularly emphasizing parallelism. Unlike superconducting platforms, where concurrent gates may increase crosstalk noise~\cite{Arute2019}, global gates available in neutral-atom platforms significantly reduce accumulated noise. Hence, maximizing global gate usage emerged as advantageous. Furthermore, while long atom transport distances impact overall circuit execution time, the primary error contributions from atom shuttling come from atom pick-up and drop-off operations and are manageable for the qubit numbers considered here. The notably long coherence times ($T_1$, $T_2$) of neutral-atom architectures~\cite{Bluvstein_2022, Bluvstein2024} adequately compensate for longer circuit execution time caused by delays introduced by atom movements. Transport distances are manageable at current scales; however, careful optimization will become critical at significantly larger scales.

We note that Quantinuum's H1-1E platform also offers dynamic connectivity; however, its gate parallelism is limited by the number of physical interaction zones. In contrast, neutral-atom hardware can simultaneously perform parallel gate operations and qubit rearrangements, with parallelism determined by the total number of atoms available. The variation in \qcrank{} infidelity observed for ion-trap processors with racetrack architectures likely stems from these hardware limitations (see diamonds near input capacity 160 in Fig.~\ref{fig:rmse_scaling}). This suggests that both the flexibility of the quantum register layout (e.g., 1D vs. 2D, number of parallelizable operations and moves) and the chosen QCrank encoding scheme jointly influence performance.
Specifically, increased parallel moves and gate operations become advantageous as the ratio $\nicefrac{n_d}{n_a}$ increases, possibly explaining the declining performance of the H1-1E processor at larger sequence lengths. Future ion-trap systems with alternative qubit layouts may yield different performance profiles, dependent on their implementation details. Finally, superconducting architectures generally exhibit poorer projected performance for this algorithm due to sparse long-range connectivity, necessitating additional native entangling gates.

While we restricted this analysis to single-zone  \dpqa\ implementations, alternative strategies such as multi-zone operation schemes~\cite{Bluvstein2024} or  rastering strategies for in-plane laser operations addressing only subsets of qubits are promising and warrant detailed exploration. 
Furthermore, as native gate fidelities improve and larger circuits are executable, \dpqa\ noise models should incorporate more detailed dynamics, including infidelities due to atom transport times, velocity profiles~\cite{Bluvstein_2022}, and atom losses.

Beyond the optimization of circuits for gate-based quantum computing, we note that the optimization of evolution protocols for analog quantum computing with neutral atoms has also been considered ~\cite{schuetz2025quantum}.

Lastly, the Z-error bias inherent in our current noise model has a particularly negative impact on the fidelity of data qubits.
Techniques such as randomized compiling~\cite{Wallman2016, Hashim_2021} 
have the potential to mitigate this issue and further enhance the effective fidelity of encoding.

\section{Methods}

\subsection{Noise model for a dynamically programmable qubit arrays }

The basic noise channels for a \dpqa~are similar as for other digital quantum computing platforms: (i) state preparation and measurement (SPAM) errors, (ii) single-qubit gate infidelities and (iii) two-qubit gate infidelities. However, the dynamic architecture of a \dpqa~introduces additional  (iv) 
errors occurring during atom rearrangement. This \emph{move-error} has to be included for each atom moving operation while executing the circuit. 
Additionally, in the considered single-zone architecture, the two-qubit  gates are implemented using a global laser pulse that illuminates all qubits, including  qubits outside the blockade radius, which are expected to idle. 
This exposure leads to photon scattering, introducing (v)  error that we refer to as \emph{CZ-spectator} error.
The local and global single-qubit gates in general have  different  fidelities. We therefore define (vi) two different single-qubit errors for \emph{local U} and \emph{global U}.
All these noise sources are taken into account in our \dpqa\ simulations and are summarized in Table \ref{tab:noisy_gates}.

In general, a noise process is described as a quantum channel $\mathcal{E} = \sum_i K_i \rho K_i^\dagger$ for a set of Kraus operators $\{K_i\}$ that map the input Hilbert space to the output Hilbert space satisfying $\sum_i K_i K_i^\dagger = \mathds{1}$. Under these conditions, $\mathcal{E}$ represents a completely positive trace preserving map on density matrices. Based on our understanding of  \dpqa\ physics, we parametrized  commonly used Pauli channels where $K_i\in\{\mathds{1}, X, Y, Z\}$ for single-qubit gates and $K_i\in\{\mathds{1}, X, Y, Z\}^{\otimes 2}$ for two-qubit gates, with $X$, $Y$, $Z$ being Pauli matrices.  More accurate noise models that incorporate leakage errors or atom loss will require more complex simulation tools and will not be considered in this work. The corresponding Pauli channels are then parametrized by probabilities $(p_x, p_y, p_z)$
\begin{equation}
\mathcal{E}(\rho) = \left(1-\sum_{i\in\{x,y,z\}}p_i\right)\rho + p_x X\rho X^\dagger + p_y Y\rho Y^\dagger + p_z Z\rho Z^\dagger
\end{equation}
for single-qubit gates and its extension for two-qubit gates $\mathcal{E}(\rho) = \sum_{P \in \mathcal{P}_2} p_P P \rho P$
where \( \mathcal{P}_2 = \{\mathds{1}, X, Y, Z\}^{\otimes 2} \) represents all possible two-qubit Pauli operations, and \( p_P \) are their respective probabilities. A special case of Pauli channel is the depolarizing channel where $p_x = p_y = p_z \equiv p$. It corresponds to the case where the qubit  decoheres to the completely mixed state with probability $p$, and with $1-p$, the qubit remains in its assigned state.

The noise model parameters in Table~\ref{tab:noisy_gates} represent reasonable approximations of current \dpqa\ capabilities~\cite{Bluvstein_2022, Evered2023, rodriguez2024experimentaldemonstrationlogicalmagic}, sufficient for evaluating algorithm performance and compilation strategies rather than capturing precise hardware noise. The ongoing progress in neutral-atom quantum computing will likely necessitate updates to this noise model.

\subsection{\qcrank\  accuracy measure}

The objective of the \qcrank\ protocol is to write a sequence of real numbers to the QPU and read them back after some optional computation.
 The natural measure of accuracy is the RMSE, computed between the measured sequence and the ground truth input. By issuing a number of shots that scale with $2^{n_a}$, we ensure that the statistical error remains constant regardless of the sequence length.

%TTTTTTTTTTTTTTTTTTTTTTTTTTTTTTTTTTTTTTTTTTTTT
\begin{table}[htbp]
\vspace{-6pt}
\begin{center}
  \caption{All \qcrank\ configurations}
  \vspace{-6pt}
\begin{tabular}{cccccc}
\hline 
\textbf{total qubits} &$\boldsymbol{n_a}$ \rule{0pt}{8pt}& $\boldsymbol{n_d}$  &  \textbf{num CZ} \dag & \textbf{ num shots} & \textbf{ IBM CZ}\ddag \\
\hline
6 &3 &3 & 24 & 25k & 66\\
9 &3 & 6 &  48 & 25k & 168\\
12 &3 & 9 &  72 & 25k & 262 \\
15 &3 & 12 &   96 &25k & 431\\
12 &4 & 8 &  128 & 50k & 546\\
10 & 5 & 5 & 160 & 100k  & 718 \\
16 &4 & 12 &  192 & 50k & 971\\
20 & 4 & 16 & 256 & 50k & 1505 \\
15 & 5 & 10 & 320& 100k &  1514\\
\hline
\end{tabular}\\
\label{tab:qcrank_size}
\dag ) also \qcrank\ input sequence length (number of stored real-valued data) \rule{0pt}{8pt}\\
\ddag ) num CZ after transpilation for IBM Fez, varies with calibration
\end{center}
\vspace{-8pt}
\end{table}
%TTTTTTTTTTTTTTTTTTTTTTTTTTTTTTTTTTTTTTTTTTTTT

\begin{figure}[htb]
\centering
\includegraphics[width=0.8\columnwidth]{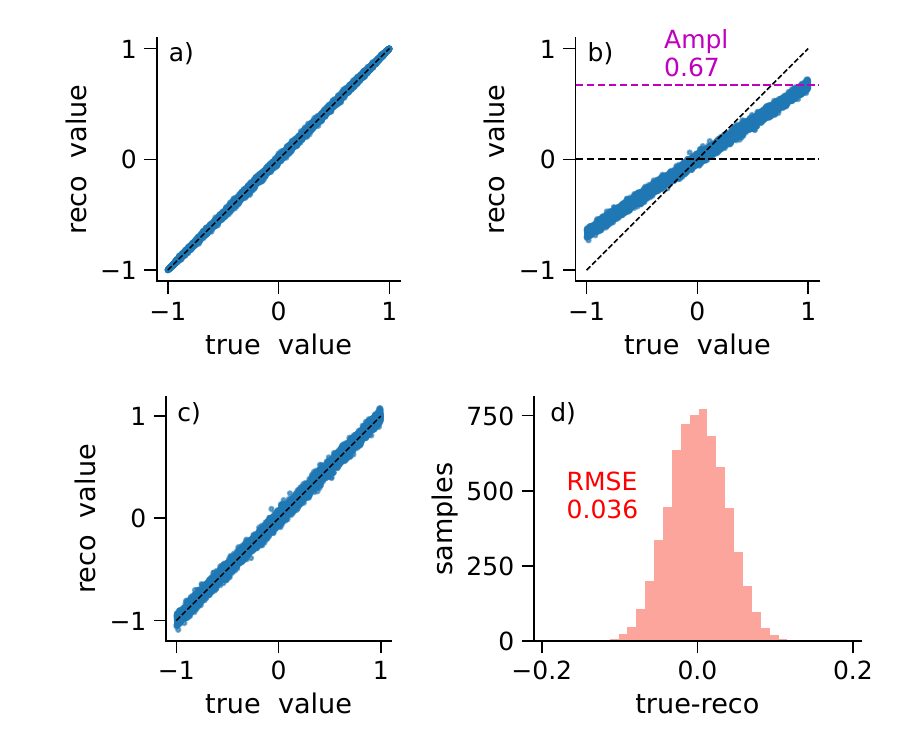}
\vspace{-12pt}
\caption{\textbf{Simulated residual for encoding of 128 
 real value sequence using \qcrank\ 4+8.} 
(a) Ideal simulator. Dashed diagonal line denotes perfect data recovery.
(b) Noisy \dpqa\ simulator output. Dynamic range is reduced to 67\% due to noise. (c) Noisy output is scaled in post-processing to recover the full dynamic range. (d)  Histogrammed residual for data from c).
}
\vspace{-12pt}
\label{fig:2D_RMSE}
\end{figure}

For statistical stability, we generate about 50 random sequences per \qcrank\ configurations, for the lengths  listed in Table~\ref{tab:qcrank_size}. As an illustration, Fig.~\ref{fig:2D_RMSE}(a) shows the correlation between reconstructed and input values for an ideal \qcrank\ simulation. The width of the ellipse provides the baseline for the magnitude of purely statistical noise, which is very small due to the large number of issued shots. Panel (b) shows  the same correlation  when a noisy \dpqa\ simulator is used with the same number of shots. 
The dynamic range of reconstructed values reduces by a factor of 0.67, the correlation tilts and fattens. In post-processing, we allow for scaling the reconstructed values by a single common factor for a given \qcrank\ setting, recovering the full dynamic range of [-1, 1]. This leads to the correlation shown in panel (c). The RMSE of \qcrank\ is defined as the width of the corrected ellipse, which we compute by building a histogram of the residuals (panel (d)) and calculating their standard deviation. We repeat this accuracy characterization for every \qcrank\ configuration.

We consider adjusting a single parameter in post-processing per configuration acceptable, as it mirrors the process of calibrating a QPU. In a real-world scenario, one would run two same input size \qcrank\ jobs on a QPU sequentially. The first job, with a known random uniform input, serves as calibration and determines the calibration constant. The post-processing of the second job, potentially with an unknown input, would then blindly use the calibration information from the previous run.

\subsection{Effect of noise model  on \qcrank\ accuracy}

The \dpqa\ noise model has  strong  $Z$-error component -- the most relevant noise source such as global CZ, CZ-spectator, and atom moves contain predominantly $Z$-Pauli errors.
The highly structured \qcrank\ encoding circuit (see Fig.~\ref{fig:qcrank2+4}) enables analysis of error propagation and informed selection of compilation strategies to reduce this Z-noise impact. $ Z$ errors on address qubits propagate through CZ gates to the final measurement and thus do not compromise data encoding quality. This underlies our strategy of moving address qubits, minimizing the effect of move-induced errors.

 In contrast, the $Z$ errors on the data qubits has a detrimental effect. $Z$ error propagates through $R_y(\theta)$ gates flipping them to $R_y(-\theta)$, eventually converting them to $X$ errors after the final Hadamard layer. We analyze the effect of this by recalling that \qcrank\ ~\cite{qcrank-nature} encodes the data $\alpha_{ij}$, where $i \in[0, 2^{n_a}-1]$, $j \in [0, n_d-1]$, into a quantum state $\ket{\Psi(\vec{\alpha})}=\sum_{i}\ket{i}_a\otimes_j \left(\cos{\alpha_{ij}}\ket{0}_j+\sin{\alpha_{ij}}\ket{1}_j\right)$, with the data related to $R_y$ rotations angles $\theta$ by the Walsh-Hadamard transform ($W$) over address index $ \theta_{ij}=\sum_{i'}W_{ii'}\alpha_{i'j}$.

We can now conclude that the final $X$ error severely obfuscates the data stored. The effect of $R_y(-\theta)$ angle flip is limited as it only affects $1/2^{n_a}$ fraction of angles participating in the $\alpha_{ij}$ through the Walsh-Hadamard transform. In contrast, $X$ Pauli error on the data qubits propagates through CZ gates by copying the $Z$ error to address qubits, with no effect. It still flips  $R_y(\theta)$ to $R_y(-\theta)$, but the effect on the data is limited by the $1/2^{n_a}$ factor. Due to this asymmetry between the effect of errors on final data encoding, better results could be obtained by either randomized compilation that symmetrizes the noise channels on data qubits or by adjusting computational basis such that the $Z$ errors become innocuous. We leave such optimizations for future work.

%%%%%%%%%%%%%%%%%%%%%%%%%%%%%%%%%%%%%%%%%%%%%%%%%%%%%%
%%%%%%%%%%%%%%%%%%%%%%%%%%%%%%%%%%%%%%%%%%%%%%%%%%%%%%

\section*{Acknowledgments}

This research used resources of two DOE user facilities: the National Energy Research Scientific Computing Center (NERSC) located at Lawrence Berkeley National Laboratory, operated under Contract No.~DE-AC02-05CH11231, and the Oak Ridge Leadership Computing Facility, operated under Contract No.~DE-AC05-00OR22725.
This work is partially funded by NSF grants CCF-2313083. AM was supported by the U.S. Department of Energy (DOE) under Contract No. DE-AC02-05CH11231, through the Office of Advanced Scientific Computing Research Accelerated Research for Quantum Computing Program.
DBT was supported by the Harvard Quantum Initiative Postdoctoral Fellowship.

\bibliographystyle{IEEEtran}

\input{references.bbl}

\end{document}

%% file: references.bbl
% Generated by IEEEtran.bst, version: 1.14 (2015/08/26)